\documentclass[11pt,a4paper,reqno]{amsart}
\usepackage{graphicx} %

\usepackage{times} %
\usepackage{amsmath} %
\usepackage{amssymb}  %
\usepackage{amsthm}
\usepackage{latexsym}
\usepackage{amsfonts,bbm}
\usepackage{xcolor}
\usepackage{mathtools}
\usepackage{enumerate}
\usepackage{tikz}
\usepackage{graphicx,caption}
\usepackage{todonotes}
\usepackage{float}
\usepackage{nicefrac}
\usepackage[foot]{amsaddr}
\usepackage{color}
\usepackage{graphicx}

\usepackage{tikz}
\usetikzlibrary{shapes.geometric}
\usetikzlibrary{arrows.meta,arrows}

\usepackage{tabularx}
\usepackage{comment}
\usepackage{cleveref}

\usepackage[normalem]{ulem}
\usepackage{algorithm}

\allowdisplaybreaks

\usepackage{amsmath,amssymb,amsfonts}
\usepackage{latexsym}
\usepackage{mathrsfs}
\usepackage{bbm}
\usepackage{xcolor}
\usepackage{xcolor}
\usepackage{tikz}
\usepackage{url}
\usepackage{stmaryrd}
\usepackage{dsfont}
\usepackage{placeins}
\usepackage{lipsum}
\usepackage{amsfonts}
\usepackage{graphicx}
\usepackage{epstopdf}
\newcommand{\R}{\ensuremath{\mathbb R}}  
\newcommand{\N}{\ensuremath{\mathbb N}}  

\newtheorem{remark}{Remark}

\newtheorem{theorem}{Theorem}

\allowdisplaybreaks

\title{Offset-free Nonlinear MPC with Koopman-based Surrogate Models}
\author{Irene Schimperna$^{1}$}\address{$^{1}$Civil Engineering and Architecture Department, University of Pavia, Italy \\ Mail: \textsc{irene.schimperna01{@}universitadipavia.it}}
\author{Lea Bold$^{2}$} \address{$^{2}$Optimization-based Control Group, Institute of Mathematics, Technische Universität Ilmenau, Germany.\\ Mail: \textsc{\{lea.bold, karl.worthmann\}{@}tu-ilmenau.de}}
\author{Karl Worthmann$^{2}$}

\begin{document}

\begin{abstract} 
    In this paper, we design offset-free nonlinear Model Predictive Control (MPC) for surrogate models based on Extended Dynamic Mode Decomposition (EDMD). The model used for prediction in MPC is augmented with a disturbance term, that is estimated by an observer.
    If the full information about the equilibrium of the real system is not available, a reference calculator is introduced in the algorithm to compute the MPC state and input references. The control algorithm guarantees offset-free tracking of the controlled output under the assumption that the modeling errors are asymptotically constant. The effectiveness of the proposed approach is showcased with numerical simulations for two popular benchmark systems: the van-der-Pol oscillator and the four-tanks process. 
\end{abstract}

\maketitle

\section{Introduction}
\noindent In recent years, the use of data-driven models for Model Predictive Control (MPC) gained increasing popularity, due to the large availability of data and the development of new effective model structures.
A large variety of data-driven model classes was successfully employed with MPC, including nonlinear ARX models \cite{denicolao1997stabilizing}, Bayesian identification \cite{piga2019performance} and neural network models \cite{ren2022tutorial,schimperna2024robust}. 
A popular approach is based on Extended Dynamic Mode Decomposition (EDMD), an approximation technique in the Koopman framework. 
The linear Koopman operator 
encodes the behavior of an associated (nonlinear) dynamical system, but is, in general, infinite dimensional. 
By using data-driven techniques like EDMD, which learns a compression of the operator to a finite-dimensional subspace spanned by so-called observable functions, a finite-dimensional approximation of the Koopman operator can be obtained, see~\cite{williams2015data}. 
Several approaches for control systems have been introduced. A widely known method is called \textit{EDMD with control} (EDMDc), see, e.g., \cite{proctor2016dynamic, korda2018linear}, which results in a linear surrogate model. 
However, as shown in~\cite{iacob:toth:schoukens:2022}, the linear surrogate resulting from EDMDc may be insufficient to fully capture state-control couplings of the underlying nonlinear system.
An alternative are 
bilinear surrogate models, see, e.g.,
\cite{peitz:otto:rowley:2020} and the references therein.
For this method, finite-data error bounds with i.i.d.\ sampling were established in \cite{nuske2023finite,schaller2023towards}.
Here, we use the stability-oriented bilinear EDMD scheme (SafEDMD) introduced in~\cite{strasser2024safedmd}, since it handles the Koopman operator directly in discrete-time and encodes knowledge of the desired set point in the structure of the surrogate-mode leveraging proportional error bounds vanishing at the origin.

EDMD-based surrogate models were used before in data-driven MPC, see, e.g., \cite{korda2018linear}. In \cite{bold2024data}, practical asymptotic stability for the EDMD-based surrogate model was proven, i.e.\ the system is steered closer to the equilibrium until the progress potentially stagnates within a certain neighborhood of the stabilized equilibrium.
This distance depends on the quality of the model used in the predictions and leads to an offset in the error.
A common method to achieve offset-free tracking is offset-free MPC. This technique was first introduced in~\cite{pannocchia2003disturbance} for linear systems and later extended to nonlinear systems \cite{morari2012nonlinear,pannocchia2015offset}. 
In offset-free MPC, the system model is first augmented with a constant disturbance term. Then, an observer is introduced to provide an estimation of the disturbance and, if they are not measurable, of the systems states. 
The disturbance estimation is used at every time step to update the reference for the optimal control problem in MPC. 
The model, augmented with the current estimate of the disturbance, is then used for prediction in MPC. 
In \cite{son2022development}, linear offset-free MPC has been used in combination with EDMDc surrogate models, showing promising results.
A similar method is employed in \cite{chen2022offset}, where EDMDc-based offset-free MPC shows an improvement of performance in the control of a soft manipulator with respect to a normal EDMDc-based MPC.

In this paper, we propose a nonlinear offset-free control algorithm for bilinear EDMD surrogates. Unlike the general case studied in offset-free MPC, with EDMD-based surrogate models, the state of the system is considered measurable. 
Hence, an observer is designed to estimate the disturbance state only. 
Two different formulations of the MPC are proposed. In the simpler first case, the target equilibrium of the system under control is considered fully known. In the second case, only partial information about the desired set-point is available. Then, a reference calculator is included in the algorithm to compute the state and input references for the MPC controller based on the estimation of the disturbance state. 
The control algorithm is tested in two simulation examples. The first is the van-der-Pol oscillator, where the information about the equilibrium is known. The second is the four-tanks process \cite{alvarado2011comparative}, where only an approximated value of the equilibrium of the system is available. 

\textbf{Notation}. We denote with $I_n$ the $(n \times n)$-identity matrix. The norm $\|\cdot\|$ is used for the Euclidean norm on $\R^n$ and its induced matrix norm on $\R^{n\times n}$.
Given a vector $v \in \R^n$ and a matrix $Q \in \R^{n \times n}$, $\|v\|_Q^2 := v^\top Q v$. 
For a number~$d\in\N$, the abbreviation $[1:d] := \mathbb{Z} \cap [1,d]$ is used. By $C_b(\Omega)$, the space of bounded continuous functions on a set $\Omega\subset\R^n$ is denoted.  

\section{Problem formulation}\label{sec:prob_form}
\noindent Consider an unknown control system, which is assumed to be control affine, i.e.\ a system
\begin{align}\label{eq:system}
    \begin{split}
        x^+ &= F(x, u) = G_0(x) + \sum\limits_{i=1}^m u^{(i)} G_i(x)
    \end{split}
\end{align}
with current and successor state $x, x^+ \in \R^n$, control $u = [u^{(1)} \ \dots \ u^{(m)}]^\top \in \mathbb{U}$, where $\mathbb{U} \subset \R^m$ is a compact and convex set.
If a continuous-time control-affine system is considered, a corresponding discrete-time sampled-data system with zero-order hold and time step $\Delta t > 0$ can be derived using the control function $u(s) \equiv u \in \mathbb{U}$ on $[0,\Delta t)$ approximately corresponding to a system of the form~\eqref{eq:system} omitting second-order terms in~$\Delta t^2$, see, e.g., \cite{nesic2004framework} or \cite[Remark~4.1]{bold2024kernel} for details.
The objective of this paper is to design an offset-free MPC algorithm based on an offline-learned data-driven surrogate model of the system \eqref{eq:system}, denoted by
\begin{align*}
    x^+ = \widehat{F}(x, u),
\end{align*}
obtained using Extended Dynamic Mode Decomposition (EDMD). 

\section{Extended Dynamic Mode Decomposition}\label{sec:EDMD}
\noindent In this section, we give a brief introduction to EDMD, which will be used in the following to obtain a data-driven model of the system under control. 

\subsection{EDMD for autonomous systems}\label{subsec:EDMD:autonomous}
Consider an autonomous discrete-time system given by 
\begin{align}\label{eq: dyn syst}
    x^+ = F(x)
\end{align}
with a map~$F:\R^n \rightarrow \R^n$.  
Now, consider system~\eqref{eq: dyn syst} on the compact and non-empty set $\Omega \subset \R^n$. 
The Koopman operator~$\mathcal{K}:C_b(F(\Omega)) \rightarrow C_b(\Omega)$, associated to system~\eqref{eq: dyn syst}, is defined by the identity 
\begin{align*}
    (\mathcal{K} \varphi)(x) = \varphi(F(x)) \qquad \forall \ x \in \R^n, \ \varphi \in C_b(F(\Omega)),
\end{align*}
where the functions~$\varphi$ are called observables in the following.
The Koopman operator is linear and bounded, but infinite dimensional. 
Let $\mathcal{D} = \{\psi_j \in C_b(\R^n):\ \ j \in [0:M]\}$ be a dictionary of finitely many observable functions, and $\mathbb{V} = \operatorname{span}\{\mathcal{D}\}$. We use EDMD to compute a finite-dimensional data-based approximation of the compression~$P_{\mathbb{V}} \mathcal{K}|_{\mathbb{V}}$ of the Koopman operators action restricted to the finite-dimensional subspace~$\mathbb{V}$, where $P_{\mathbb{V}}$ denotes the projection on $\mathbb{V}$.
Let $\Psi = (\psi_0, \psi_1, \dots, \psi_M)^\top$ be the stacked vector of the observables with $\Psi: \R^n \rightarrow \R^{M+1}$ and set $\psi_0(x) \equiv 1$. 
Then, consider a set of $d$ i.i.d.\ sampled data points~$x_i\in \Omega$, $i \in [1:d]$, and compute the values $y_i = F(x_i)$. 
Each data point is lifted onto the set~$\Psi(\Omega) = \{z \in \R^{M+1}|\ \exists\ x \in \Omega \text{ with } \Psi(x) = z\}$ embedded in $\mathbb{R}^{M+1}$.
In the following, we assume that the coordinate functions are included as observables, i.e.\ $\psi_j(x) = e_j^\top x$ for $j \in [1:n]$, where $e_j$ denotes the $j$-th unit vector of $\R^n$, as proposed in \cite{mauroy2019koopman}. 
The data points are arranged in the data matrices
\begin{align*}
    \Psi_X := [\Psi(x_1) \ \dots \ \Psi(x_d)] \text{ and } \Psi_Y:= [\Psi(y_1) \ \dots \ \Psi(y_d)].
\end{align*}
The approximation~$\widehat{K}$ of $P_{\mathbb{V}} \mathcal{K}|_{\mathbb{V}}$ is then given by the solution of the linear regression problem 
\begin{align*}
    \widehat{K} = \operatorname{argmin}\limits_{K \in \R^{(M+1) \times (M+1)}}\|K\Psi_X - \Psi_Y\|,
\end{align*}
which is explicitly given by
\begin{align*}
    \widehat{K} = C^{-1} A \quad \text{ with } \quad C = \Psi_X\Psi_X^\top,\ A = \Psi_X\Psi_Y^\top. 
\end{align*}
The EDMD-based surrogate model then reads $x^+ = P_{\Omega} \widehat{K}\Psi(x)$,
where $P_{\Omega}$ is the coordinate projection onto the state space.
Since the coordinate functions are observables, the coordinate projection can be performed using
    \begin{align*} 
    x^+ = \left[ \begin{array}{ccc}
            0_{n \times 1} & I_n & 0_{n \times (M - n)}
    \end{array} \right] \widehat{K}\Psi(x). 
\end{align*}

\subsection{EDMD for control-affine systems}\label{subsec:EDMD:control}

Next, EDMD is extended to control-affine discrete-time systems~\eqref{eq:system}. 
For the bilinear method as in \cite{peitz:otto:rowley:2020}, we use the fact that the Koopman operator approximately preserves control affinity. 
So, for a control value~$u \in \mathbb{U}$, an approximation of the Koopman operator~$\widehat{K}_u$, corresponding to the constant control function $u(t) \equiv u$ and, thus, to an autonomous dynamical system, can be constructed by 
\begin{align}\label{eq:Koopman:u}
    \widehat{K}_u = \widehat{K}_{u_0} + \sum_{i = 1}^{m} \lambda_i \cdot \left(\widehat{K}_{u_i} - \widehat{K}_{u_0}\right),
\end{align}
where the fixed control input~$u_0 = 0$ and the inputs~$u_i \in \mathbb{U}$, $i\in [1:m]$, provide a basis of the input space~$\mathbb{R}^m$. 
The matrices $\widehat{K}_{u_i}$ are then generated using the EDMD method in the autonomous case as introduced in Section~\ref{subsec:EDMD:autonomous}. 
The coefficients $\lambda_i$ solve the linear system of equations $\sum_{i = 1}^{m} \lambda_i u_i = u$. 
Hence, if the inputs are chosen as unit vectors of $\R^m$, that is, $u_i = e_i$, then $\lambda_i = u^{(i)}$ holds. 
The matrix~$\widehat{K}_{u}$ given by~\eqref{eq:Koopman:u} denotes the matrix representation of the approximated Koopman operator for system~\eqref{eq:system} with a constant control input~$u$. 
Notice that for the approximation of the operators $\widehat{K}_{u}$ we do not need identical samples~$x_1, \dots, x_d$ for the regression. It is convenient to have $m+1$ data sets~$x_1^{[i]}, \dots, x_d^{[i]}$ for $i \in [0:m]$. 
Using this procedure, the EDMD-based surrogate model can then be set up as
\begin{align}\label{eq:propagation_Koopman}
    x(k + 1) = P_\Omega\widehat{K}_{u(k)} \psi(x(k)),
\end{align}
where we introduced the time index~$k \in \mathbb{N}$ to indicate the prediction using the bilinear EDMD surrogate~\eqref{eq:propagation_Koopman} for control sequences $(u(k))_{k=0}^{N-1}$, $N \in \mathbb{N}$. To this end, an initial condition $x(0) = \hat{x}$ is added.

For systems of the form~\eqref{eq:system}, that have an equilibrium in the origin, i.e.\ $F(0, 0) = 0$, it is possible to exploit the structure of the observables, in particular the fact that $\psi_0 (x) \equiv 1$, to enforce a certain structure of the Koopman matrices. This is done, e.g., in SafEDMD proposed in~\cite{strasser2024safedmd}. 
More precisely, the Koopman matrices become 
\begin{equation}\label{eq:SafEDMD}
    \widehat{K}_0 = \begin{bmatrix}
        1 & 0 \\
        0 & A
    \end{bmatrix}, \qquad \widehat{K}_i = \begin{bmatrix}
        1 & 0 \\
        b_{i} & B_i
    \end{bmatrix}
\end{equation}
with $\widehat{K}_i \in \R^{(M + 1) \times (M + 1)}$ for all $i \in [0 : m]$ and $A, B_i \in \R^{M \times M}$ and $b_{i} \in \R^M$. This ensures that the (controlled) equilibrium is preserved in the data-driven surrogate model.

\section{Offset-free MPC}
\noindent In this section, we present the offset-free MPC algorithm.
Therein, the model is augmented by a disturbance term, which is estimated online by an observer and used, together with the model, in the optimization step of the MPC scheme. 
A specific observer structure is proposed for EDMD surrogate models, where the state is measurable and only the estimation of the disturbance term is required.
The offset-free MPC algorithm can handle both, i.e., the following two distinct cases:
\begin{enumerate}
    \item [(ke)] \textit{Known Equilibrium}: full information about the system equilibrium is available.
    \item [(ue)] \textit{Unknown equilibrium}: the case when only partial knowledge about the equilibrium is available.
\end{enumerate}
In the following, we will first present the offset-free MPC algorithm for the known equilibrium case~(ke), and then we will introduce the modifications that are needed to manage the case~(ue), where only a partial information about the equilibrium is available.

\subsection{Case of known system equilibrium (ke)}

The objective of the MPC is to track without offset a constant reference, that is an equilibrium of the real system. 
In some cases the physics of the system allows us to know the full information about the equilibrium $(\bar{x}, \bar{u})$ of system~\eqref{eq:system}, that we want to track. 
In the following, it is assumed that $\bar{u}$ is in the interior of $\mathbb{U}$.
The control objective for the closed-loop system is
\begin{align*}
    \lim_{k \to \infty} x(k) = \bar{x}, \qquad \lim_{k \to \infty} u(k) = \bar{u}.
\end{align*}

In offset-free MPC, the model used in the optimal control problem is augmented with a term $\hat{d} \in \R^n$ to compensate for the presence of modeling errors. In particular, this term is added to the state equation of the model, and is estimated by an observer with the following equations
\begin{align}\label{eq:observer}
\begin{split}
    \tilde{x}(k) &= \widehat{F}(\hat{x}(k-1), u(k-1)) + \hat{d}(k-1), \\
    \hat{x}(k) &= x(k), \\
    \hat{d}(k) &= \hat{d}(k-1) + (\hat{x}(k) - \tilde{x}(k)).
\end{split}
\end{align}
The observer includes a state $\hat{x} \in \R^n$ that is used to keep in memory the previous value of the state of the system, but is not used by the MPC, and a term $\tilde{x} \in \R^n$ that is the one-step prediction of model corrected by $\hat{d}$ starting from the previous value of the system state. The output of the observer is $\hat{d}$, that is used to update the MPC prediction model. If the system state is unmeasurable, a state and disturbance observer can be employed, as shown, e.g., in \cite{morari2012nonlinear}.

The MPC is designed based on the augmented model with cost function~$\ell(x, u): \R^n \times \R^m \rightarrow \R_{\ge 0}$. 
At time step $k$, the MPC algorithm solves the following optimal control problem
\begin{align}\label{eq:OCP}
\begin{split}
    \min_{u = u(k+\cdot)} \ & \sum_{i = 0}^{N-1}  \ell (x(k + i) - \bar{x}, u(k + i) - \bar{u}) \\
    \text{s.t.} \ &  x(k+i+1) = \widehat{F} (x(k+i), u(k+i)) + \hat{d}(k) \\
    & u(k+i) \in \mathbb{U} \qquad\text{for all $i \in [0:N-1]$},
\end{split}
\end{align}
given the measured initial value~$x(k)$, where 
$N \in \mathbb{N}$ is the prediction horizon.
Note that the disturbance estimation from the observer is added to the state equations in the MPC optimal control problem~\eqref{eq:OCP}, and it is kept constant along the horizon. 
The MPC control determines the control value~$u(k)$. It is given by
$u(k) = \mu_N(x(k), \hat{d}(k)) = u^*(k)$,
where $u^*(k)$ is the first element of the optimal input sequence.
\begin{remark}
    Our MPC formulation does not consider a terminal cost or a terminal constraint. Hence, a sufficiently long prediction horizon should be used to guarantee recursive feasibility and stability of the MPC closed loop, see~\cite{boccia2014stability}.
\end{remark}

\begin{remark}
    In the description of offset-free MPC we have considered a system model in the form of \eqref{eq:propagation_Koopman}, where a projection operation is performed at each prediction step. The projection step improves the quality of the predictions, in particular with long prediction horizon, see, e.g., \cite{mauroy2019koopman} and~\cite[Remark 20]{nuske2023finite}. 
    However, it is also possible to implement the offset-free MPC in the lifted space, without including a projection operation at each prediction step. In this case the disturbance state has the dimension of the lifted space, i.e. $\hat{d} \in \R^{M + 1}$, and the prediction model used in the MPC algorithm has the form
    \begin{align*}
        z(k) &= \psi(x(k)), \\
        z(k+i+1) &= \widehat{K}_{u(k)} z(k+i) + \hat{d}(k), \quad i \in [0:N-1].
    \end{align*}
    The observer has the same structure of \eqref{eq:observer}, but is designed in the lifted space.
\end{remark}

\subsection{Case of unknown system equilibrium}

In some systems only partial information about the equilibrium is available. 
An example is when we only know the values of a subset of the states at the equilibrium.
The controlled output is $y_\mathrm{c}$ that is a function of the state
\begin{align*}
    y_\mathrm{c} = r(x)
\end{align*}
and $\bar{y}_c$ is the desired setpoint for $y_\mathrm{c}$. Then $\bar{y}_c$ is the only available information about the equilibrium, that we want to reach.
The control objective for the closed-loop is to reach an equilibrium such that
\begin{align*}
    \lim_{k \to \infty} y_\mathrm{c}(k) = \bar{y}_\mathrm{c}.
\end{align*}
To derive the full information about the equilibrium to be used in the MPC cost function, a reference calculator is introduced in the algorithm.
The reference calculator solves the following optimization problem
\begin{subequations} \label{eq:ref}
    \begin{align}
    \min_{x, u} \quad & \ell_\mathrm{r}(x, u)  \label{eq:ref-1} \\
    \text{s.t.} \quad & x = \widehat{F}(x, u) + \hat{d} \label{eq:ref-2} \\
    & r(x) = \bar{y}_\mathrm{c} \label{eq:ref-3} \\
    & u \in \mathbb{U},  \label{eq:ref-4}
\end{align}
\end{subequations}
where $\ell_\mathrm{r}:\R^n \times \R^m \rightarrow \R_{\ge 0}$ is a steady state cost function.
It is assumed that \eqref{eq:ref} is feasible and that its (unique) solution is denoted by $(\bar{x}, \bar{u})$ and represents the reference to be used by the MPC.
If \eqref{eq:ref-2}-\eqref{eq:ref-3} have a unique solution, then it is not needed to implement the reference calculation as an optimization problem, but it is sufficient to compute the unique solution of the system of equations.
The MPC references are updated at every time step depending on the current value of $\hat{d}$.
All the other parts of the algorithm are implemented as described in the previous subsection. 
The complete algorithm is summarized in Algorithm~\ref{alg:ofMPC}.
\begin{algorithm}[htb]
    \caption{Offset-free MPC}\label{alg:ofMPC}
    \raggedright
    \smallskip\hrule
    \smallskip
    {\it Input:} Horizon $N \in \N$, stage costs $\ell: \R^n \times \R^m \rightarrow \R_{\ge 0}$, data-driven model~$\widehat{F}$, input constraints~$\mathbb{U}$, steady state cost function $\ell_\mathrm{r}:\R^n \times \R^m \rightarrow \R_{\ge 0}$, controlled output function~$r$, controlled output set point $\bar{y}_\mathrm{c}$.
    \smallskip\hrule
    \medskip
    \textit{Initialization}: Set $k = 0$ and  $\hat{d}(0) = 0$. \\[2mm]
    \noindent\textit{(1)} Measure current state $x$ and set $x(k) = x$.\\[1mm]
    \noindent\textit{(2)} Solve~\eqref{eq:ref} to obtain the reference~$(\bar{x}, \bar{u})$. \\[1mm]
    \noindent\textit{(3)} Solve the optimal control problem~\eqref{eq:OCP} to obtain the optimal control sequence~$u^*(k + i)_{i = 0}^{N - 1}\subset \mathbb{U}$. \\[1mm]
    \noindent\textit{(4)} Set the feedback law~$\mu_N(x(k), \hat{d}(k)) = u^*({k})$ and shift $k = k + 1$. \\[1mm]
    \noindent\textit{(5)} Update the disturbance~$\hat{d}(k)$ with equation~\eqref{eq:observer} and go to (1).
    \smallskip\hrule
\end{algorithm}

Now, we recall the theoretical results available in the literature for offset-free MPC, which also apply with EDMD-based surrogate models.
First, define the modeling error as
\begin{align*}
    w(k) := F(x(k), u(k)) - \widehat{F} (x(k), u(k)).
\end{align*}
Then, the following theorem follows from Theorems 14 and 15 in \cite{pannocchia2015offset}.
\begin{theorem}
    Assume that $w$ is bounded and asymptotically constant, i.e. there exists $\bar{w}$ such that
    \begin{align*}
        \lim_{k \to \infty} w(k) = \bar{w}.
    \end{align*}
    Then $\lim_{k \to \infty} x(k) - \tilde{x}(k) = 0$ and $\lim_{k \to \infty} \hat{d}(k) = \bar{w}$, where $\tilde{x}$ is defined in \eqref{eq:observer}. Moreover, if the closed-loop system reaches an equilibrium with input $u_\infty$ and state $x_\infty$, then $r(x_\infty) = \bar{y}_c$.
\end{theorem}

\begin{remark}
    The theorem is formulated in considering the system equilibrium unknown. The case with the known system equilibrium can be seen as a particular case.
\end{remark}

\begin{remark}
    The error bounds of the EDMD model has been exploited in \cite{bold2024data} to prove practical asymptotic stability of EDMD-based MPC. Practical asymptotic stability means that the tracking error reaches a neighborhood of the origin, but in general it does not go to zero in view of the presence of modeling errors. The proposed offset-free MPC algorithm overcomes this issue, allowing the system to asymptotically reach the reference with zero error.
\end{remark}

\section{Simulation results}

\subsection{Van-der-Pol oscillator}
For our first example, we consider the non-linear control-affine van-der-Pol oscillator
\begin{align}\label{eq:vdP}
    \dot{x}(t) = \binom{x_2(t)}{\nu (1 - x_1(t))^2x_2(t) - x_1(t) + u}
\end{align}
with parameter $\nu = 0.1$. We consider the ODE~\eqref{eq:vdP} as a sampled data system as in Section~\ref{sec:prob_form}, where the integrals are numerically solved using the classic Runge-Kutta method of fourth order and step size~$\Delta t = 0.05$. This results in a discrete-time system  
\begin{align}\label{eq:vdP:discretized}
    x(k + 1) = F(x(k), u(k)), \qquad x(0) = x^0
\end{align}
which serves as ground truth. 
We point out that the system \eqref{eq:vdP:discretized} is not control affine as assumed in \eqref{eq:system}, but the offset-free MPC is still able to exactly stabilize
it, as shown in the following.
For the approximation of the Koopman operator, EDMD as described in Section~\ref{subsec:EDMD:control} is used. The set of $d = 1000$ data points is sampled on the set~$\Omega = [-2, 2]^2$. We have used a large number of data points to ensure a high prediction accuracy of the model, but similar results are obtained considering $d=100$.
The dictionary of observables is chosen to be $\mathcal{D} = \{\psi(x) = x_1^p x_2^q: \ p, q \in [0:3] \text{ and } p + q \le 3\}$, so it contains all monomial functions of the state variables up to degree 3. Other possible choices for the observables functions rely on data-driven methods, such as the use of kernels (\cite{bold2024kernel}) or neural networks (\cite{yeung2019learning}).

The objective of MPC is to reach the equilibrium point in the origin of the system, that is, $(\bar{x}, \bar{u}) = (0,  0)$. In this example the value of the equilibrium can be easily deduced from the physical meaning of the state and input variables.
The MPC cost function is chosen to be the quadratic function~$\ell(x, u) = \|x\|_Q^2 + \|u\|_R^2$ with weighting matrices~$Q = I_2$ and $R = 10^{-2}$, subject to $-2 \le u(k) \le 2$. For our simulations, the horizon $N = 50$ is set.
\begin{figure}
    \centering
    \includegraphics[width=0.5\linewidth]{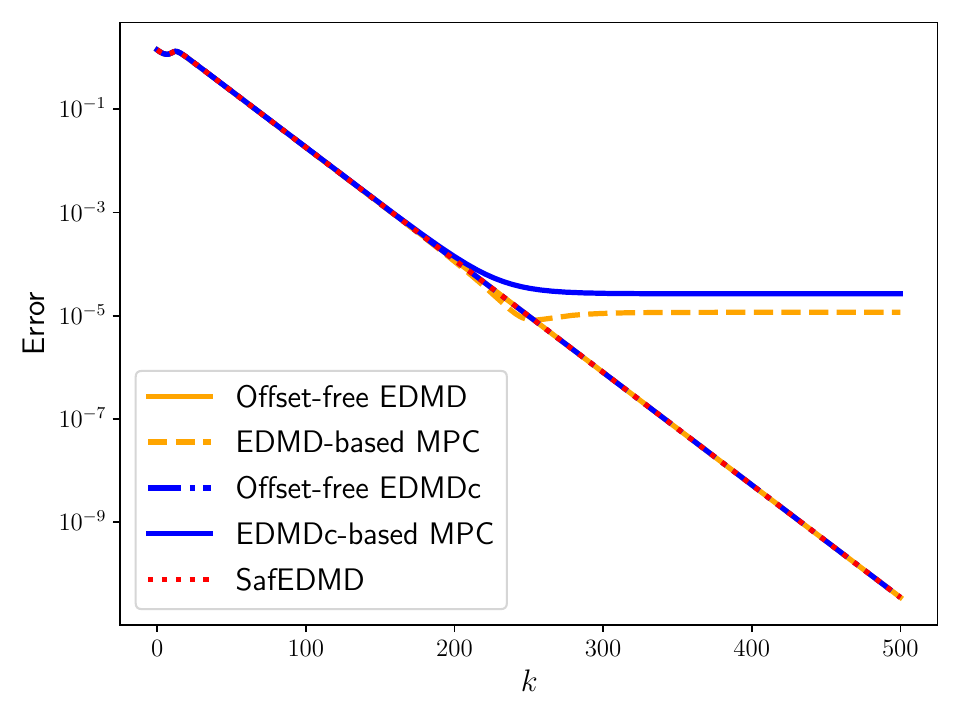}
    \caption{Norm~$\|x(k)\|$ of the closed-loop solution of the van-der-Pol oscillator~\eqref{eq:vdP:discretized} for bilinear EDMD-based and EDMDc-based MPC and offset-free MPC, and for SafEDMD-based MPC.}
    \label{fig:vanderPol:EDMDvsEDMDc}
\end{figure}
In Fig.~\ref{fig:vanderPol:EDMDvsEDMDc} 
the errors in closed-loop simulations with initial state $x^0 = (1.0, 1.0)^\top$ are reported. 
The bilinear EDMD-based offset-free MPC and a standard bilinear EDMD-based MPC 
are compared to (offset-free) EDMDc-based MPC. In addition, a SafEDMD-based MPC algorithm is performed.
It can be seen that standard MPC can only provide practical asymptotic stability (see \cite{bold2024data}), and the error stagnates at a value close to $10^{-5}$, which also occurs in the EDMDc-based MPC case with a slightly larger offset. Instead, offset-free MPC, for both EDMD- and EDMDc-based algorithms, 
provides exponential convergence towards the origin, and the closed-loop behavior of the two algorithms is almost the same. 
The same precision is reached by using SafEDMD-based MPC without using the offset-free technique. This is due to the structure of the Koopman matrices in~\eqref{eq:SafEDMD}, 
which preserves the equilibrium of the system.

\subsection{Four tanks system}
As a second example we consider the four tanks system described by \cite{alvarado2011comparative} and characterized by the following differential equations
\begin{align}\label{eq:fourTanks}
\begin{split}
    \Dot{h}_1 &= - \frac{a_1}{S} \sqrt{2gh_1} + \frac{a_3}{S} \sqrt{2gh_3} + \frac{\gamma_a}{S} q_a, \\
    \Dot{h}_2 &= - \frac{a_2}{S} \sqrt{2gh_2} + \frac{a_4}{S} \sqrt{2gh_4} + \frac{\gamma_b}{S} q_b, \\
    \Dot{h}_3 &= - \frac{a_3}{S} \sqrt{2gh_3} + \frac{1 - \gamma_b}{S} q_b, \\
    \Dot{h}_4 &= - \frac{a_4}{S} \sqrt{2gh_4} + \frac{1 - \gamma_a}{S} q_a.
\end{split}
\end{align}
The state of the system is given by the levels in the four tanks, i.e. $x = (h_1, h_2, h_3, h_4)^\top \in \R^4$, while the control variables are the flows in the two valves, i.e. $u = (q_a, q_b)^\top \in \R^2$. The numerical values of the system parameters can be found in \cite{alvarado2011comparative}, and the sampling time for the system is $\Delta t = 25s$. 
As in the previous example, the classic Runge-Kutta method of fourth order is used to numerically solve the integrals.
For this system only an approximation $(\tilde{x}, \tilde{u})$ of the equilibrium is available and is given by $\tilde{x} = (0.65, 0.66, 0.65, 0.66)^\top m$ and $\tilde{u} = (1.63, 2.0)^\top m^3/h$. The objective of the control is to drive the controlled output $r(x) = (x_1, x_2)^\top$ to the value $\bar{y}_\mathrm{c} = (\tilde{x}_1, \tilde{x}_2)$ for input constraints~$\mathbb{U} = [0, 3.26]m^3/h \times [0, 4]m^3/h$.
First, we carry out some simulations assuming that the full equilibrium information is available. These simulations are denoted with ``known equilibrium” (ke) in the following.
To do so, the real value of the system equilibrium $(\bar{x}, \bar{u})$ associated to $\bar{y}_\mathrm{c}$ is calculated using the system equations, which leads to $\bar{x} = (0.65, 0.66, 0.6417, 0.6882)^\top m$ and $\bar{u} = (1.666, 1.974)^\top m^3/h$.
Note that this is in general not possible in a realistic setting, where only data from the system are available.
For the derivation of the model, the states and inputs are shifted so that the equilibrium $(\bar{x}, \bar{u})$ corresponds to the origin, and the inputs are scaled using their maximum value, so that each component of the shifted and scaled input lies in the set $[-2, 2]$.
As observables, we used monomials up to degree 2, resulting in $M = 14$ functions, and we considered $d = 1000$ data points for each approximation of the Koopman operator. The state data are randomly sampled with a uniform distribution in the set $\Omega = [0.2, 1.36]^2 \times [0.2, 1.30]^2$.
Then some simulations are performed without using the information about the real value of the system equilibrium and are denoted by ``unknown equilibrium” (ue). In this case, the model is obtained by shifting the states and inputs so that the approximated equilibrium $(\tilde{x}, \tilde{u})$ corresponds to the origin. In both the known equilibrium case and the unknown equilibrium case, the offset-free MPC is compared with a standard EDMD-based MPC.
For all closed-loop simulations, the initial state of the system is $x^0 = (1.0, 1.0, 1.0, 1.0)^\top$.
The cost of the MPC is $\ell(x, u) = \|x - \bar{x}\|_Q^2 + \|u - \bar{u}\|_R^2$, where $Q = I_4$, $R = 10^{-4} \cdot I_2$, and the prediction horizon is set to $N = 50$. In the offset-free MPC in the unknown equilibrium case, the reference calculator solves \eqref{eq:ref-2}-\eqref{eq:ref-3} using Newton's method at every time step to compute the reference $(\bar{x}, \bar{u})$ for the MPC. In the simulation with standard MPC and unknown equilibrium, the approximated equilibrium $(\tilde{x}, \tilde{u})$ is used in the MPC cost instead of $(\bar{x}, \bar{u})$.
\begin{figure}
    \centering
    \includegraphics[width=0.5\linewidth]{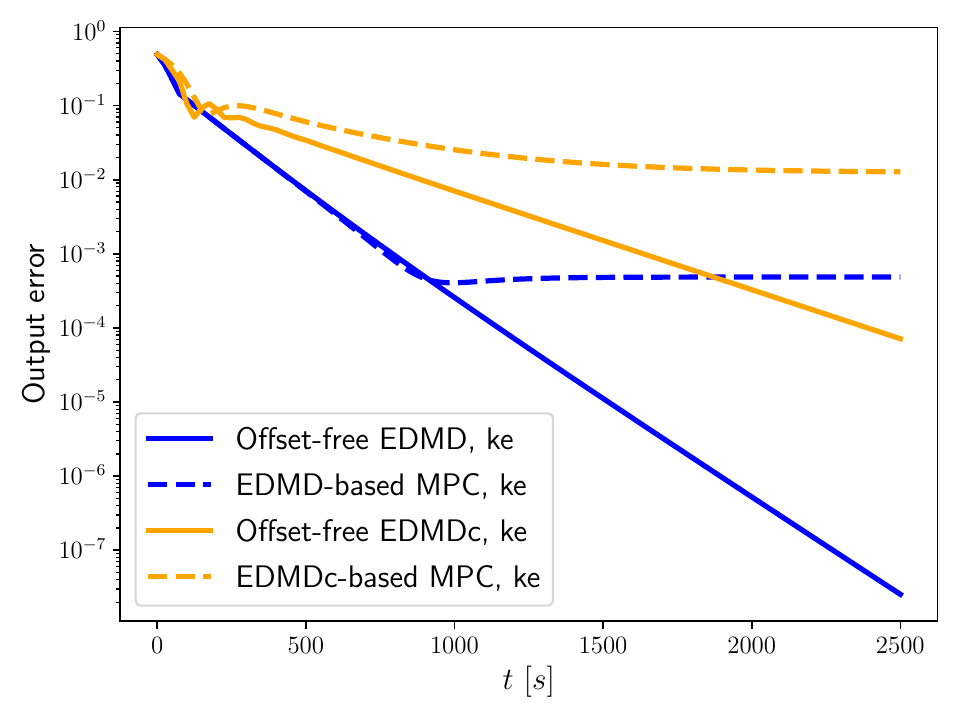}
    \caption{Output error $\|r(x) - \bar{y}_\mathrm{c} \|$ of System~\eqref{eq:fourTanks} for EDMD-based MPC and offset-free MPC comparing EDMDc and the bilinear approach for a known equilibrium.}
    \label{fig:fourTanks_errors:EDMDc}
\end{figure}
\begin{figure}
    \centering
    \includegraphics[width=0.5\linewidth]{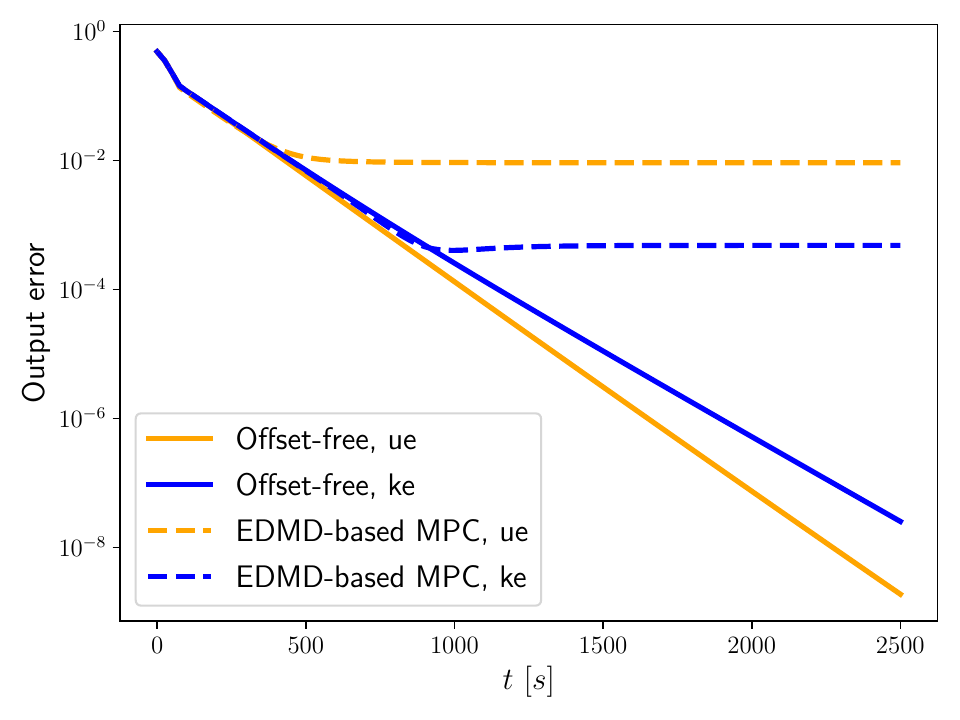}
    \caption{$\|r(x) - \bar{y}_\mathrm{c} \|$ of System~\eqref{eq:fourTanks} for EDMD-based MPC and offset-free MPC, in the cases with unknown equilibrium (ue) and known equilibrium (ke).}
    \label{fig:fourTanks_errors}
\end{figure}
Figure~\ref{fig:fourTanks_errors:EDMDc} shows the comparison of MPC and offset-free MPC based on bilinear EDMD and EDMDc in the tracking error on the controlled output for the scenario of a known equilibrium. As in the previous example, the errors of both MPC algorithms stagnate, whereas offset-free MPC leads to higher precision. In this example, the difference between bilinear EDMD and EDMDc becomes very obvious. Not only is the offset for EDMDc clearly higher when using MPC, but also the decay of the error is noticeably slower with offset-free MPC. 
In Figure~\ref{fig:fourTanks_errors}, the norm of the tracking error for EDMD-based MPC and offset-free MPC is pictured for the two cases of a known and an unknown equilibrium. 
The error is larger when the real value of the equilibrium is unknown, and an approximation is used in the MPC implementation. Steady-state error is instead not present in the simulations with the offset-free MPC, both in the known and unknown equilibrium cases.

\section{Conclusion}
\noindent
We proposed an offset-free algorithm for the control of systems modeled using EDMD. The observer is designed to estimate a disturbance term that is used to modify the prediction model of the MPC. 
To handle the case when the full information about the equilibrium is unknown, a reference calculator can be included in the closed loop to provide the state and input references for the MPC. 
The effectiveness of the proposed approach is illustrated with two simulation examples, in which the offset-free MPC provides better performance compared to the standard EDMD-based MPC. 
Future work includes the application of offset-free MPC to the kernel-EDMD (kEDMD) setting for control systems using the recently proposed kEDMD algorithm for control-affine systems allowing for flexible sampling of control-state data and, in addition, pointwise bounds on the full approximation error, see~\cite{bold2024kernel}.

\bibliographystyle{plain}
\bibliography{bibliography}

\end{document}